\documentclass[aps]{revtex4}
\usepackage{graphics}
 
\def\btau{\mbox{\boldmath $\tau$}}
\def\bpi{\mbox{\boldmath $\pi$}}
\def\br{\mbox{\boldmath $r$}}

\newcommand{\be}{\begin{equation}}
\newcommand{\ee}{\end{equation}}
\newcommand{\ba}{\begin{array}}
\newcommand{\ea}{\end{array}}

\begin{document}

\pagestyle{plain}


\title{
{\bf  Pion Mass Dependence of Nucleon Magnetic Moments } 
}
\author{
{\bf  F. L. Braghin \thanks{ braghin@if.usp.br}
,
\\ 
{\normalsize 
Instituto de F\'\i sica da Universidade de S\~ao Paulo,} \\
{\normalsize C.P. 66.318,  C.E.P. 05315-970, S\~ao Paulo, SP     
Brasil }
} }

\begin{abstract}
The relevance of the pion mass, provenient from a 
term which explicitely breaks chiral symmetry in the Lagrangian,
for nucleon magnetic moment in the frame
of the Skyrmion model in two different versions: the usual Skyrme
model and a modified one which includes a coupling to a light 
scalar meson field, the sigma $\sigma (\simeq 500-600$ MeV).
The results are compared to other calculations.
Our main motivation comes from usual extrapolations 
for values of low energy QCD observables obtained in lattices 
with large values of pion/quark masses 
toward realistic value of $m_{\pi}$.
which do not allow it.
We do a comparison with results from the Cloudy Bag Model 
 and a chiral hadronic model from chiral perturbation theory.
There are several resulting extrapolations from the region 
of large pion mass to the realistic value depending on the 
considered model for low 
energy QCD. 
\end{abstract}

\maketitle



IF-USP, IFT-UNESP: 1999/2003


\section{ Introduction}

Quantum Chromodynamics is a very intrincate theory with 
flavor and color non abelian structures
and strong coupling constants at low energies. 
There are basically two ways of extracting physical observables
from QCD. 
Lattice calculations in discretized space-time system and 
effective models with the relevant
hadronic, quark and gluon degrees of freedom considering
the more fundamental properties and symmetries.

In most of QCD lattice calculations it is not possible to obtain
full reliable results mainly because the lattices can not
be large enough and it does not provide realistic values for quark/pion
 masses 
\cite{LLT,HEMMERTWEISE}. 
Therefore it is needed 
to perform extrapolations from exact calculations in
finite lattice to the physical region
of observables to ensure reliability.
For this task one can make use of 
effective models for the description of low and intermediary energy 
QCD. 
They are based on general properties, symmetries and
their dynamical breakdowns of the believed more fundamental theory, QCD.
The non linear sigma model with pions, and extensions which gave 
origin to Chiral Perturbarion Theory (CPT), is obtained from the linear 
realization by fixing the fourth degree of freedom, the sigma field.
This is possible due to the dynamical chiral 
symmetry breaking expected to be realized in the Nambu mode. 
Recently there has been claimed that a scalar sigma particle 
($\sigma(\simeq 500-600 MeV)$  exists
 \cite{TESEME,RT98+DIGHOLE,TORNQVIST,BEDIAGA,7a}). 
However such field may develop a classical counterpart
(the
chiral condensate of the chiral dynamical breaking), 
has been neglected in many kinds of calculations including 
lattice calculations.
Therefore it becomes very interesting to study the effect of such mesonic
degree of freedom at the phenomenological and theoretical levels
like it has been done in the Skyrmion model 
and other models \cite{TESEME,7b}.
In particular in \cite{TESEME,RT98+DIGHOLE,FLBcont,7c,CR01} 
it is shown that this coupling
may lead to a redefinition of the chiral radius (which may become
a non homogeneous function), which can be different 
from the classical value
of the sigma field (or equivalently the scalar quark-anti-quark 
condensate).

The quarks masses, and also the pion masses, are expected to break 
Chiral symmetry explicitely appearing in Lagrangian terms 
\footnote{ Descriptions and origins 
for these masses are proposed 
 respectively in \cite{FLBcont}.}.
An expected relation between these variables,
of  QCD and hadronic degrees of freedom, can be seen
 through the Gell-Mann--Oakes--Renner
 relation which, to the lowest order,
 reads:
\be \label{1} \ba{ll}
\displaystyle{ 2 <\bar{q} q> m_q =  - f_{\pi}^2 m_{\pi}^2 . }
\ea
\ee
Therefore one is to consider that large pion mass is, in principle, 
equivalent
to large quark mass which break, both, chiral symmetry in the same way
\footnote{ Alternative proposal may be envisaged 
\cite{FLBcont,TIMOTEOLIMA}.}.

In this work we want to address the reliability of these 
extrapolations by  considering different large pion masses
for the calculation of nucleon static properties, 
in particular their magnetic moments,
by using two different versions of the Skyrmion model. 
The results are compared to previous calculations.
The Adelaide group \cite{LLT} used the cloudy bag model for giving
such a extrapolation for the dependence of the nucleons magnetic moments
on the pion mass. 
This has been object of attention also of 
 \cite{HEMMERTWEISE} which considered chiral perturbation theory.
It is shown that there are different possibilities of extrapolating
results from the region in which pion mass is large to the chiral limit 
depending on the model and its physical content.

\section{ The Skyrmion model and extension with scalar meson field}

\subsection{ The Skyrmion}

The Skyrmion model has been 
extensively discussed in the literature therefore we omit details about it.
The Skyrmion model  
is basically a solitonic solution of the non linear sigma model
with a non trivial topology. 
The usual non linear sigma model with the pion field
alone however does not possesses stable
solitonic solutions and a stabilizing term with higher order 
derivatives (usually fourth order) is introduced as proposed by
Skyrme long ago \cite{SKYRME} - sixth and higher order derivative terms 
may also be considered - see for example  \cite{orildo}.
These fourth and sixth order terms are obtained with chiral invariant 
couplings of the mesons rho and omega fields to the non linear sigma model
making the limit of their masses going to infinity. This is only
possible with a constant fixed pion decay constant, $f_{\pi}=\eta$, 
which, in the 
vacuum, is equal to the classical scalar field (and to the chiral radius) 
but may not be inside the hadrons
\cite{RT98+DIGHOLE}.

The winding number of the topological soliton is identified to 
the baryonic number (topologically conserved)
of the resulting soliton, which is interpreted as a baryon.
The solitonic character is responsible for the stability of such
extended body.

The simplest usual Skyrmion Lagrangian density will be considered
firstly as proposed by Skyrme \cite{SKYRME}
with the Hedgehog ansatz (given by $\bpi = \br$) for the invariant function:
$U ( \br ) = \cos(F(\br)) + i { \btau \cdot \hat{\bpi}} \; \sin(F(\br)).$
The coupling constant $e$ may be related to the pion-pion scattering
coupling  \cite{TRUONG}.
The values usually adopted for this coupling ($e \simeq 4 -6$) are also found
in large $N_c$ analysis with extrapolation to $N_c=3$ 
using sum rules of \cite{DOSCH} ($e\simeq (7.6-12)/\sqrt{N_c}$).
Its value is usually considered to be independent of $f_{\pi}$. 

\subsection{ Coupling a scalar sigma field: the Skyrmignon} 

The "natural" way of imposing a  coupling of a scalar meson field 
(such as the chiral partner of the pion) to the
Skyrmion model is to consider the linear realization of the sigma model
\cite{LSM}.
The scalar degree of freedom
may be represented by a dynamical/variable 
scalar field, $\eta(r)$, which becomes the (new) scalar
dynamical degree of freedom \cite{TESEME,RT98+DIGHOLE}.
This is equivalent to make the chiral radius dependent on the 
spatial coordinate due to the chiral symmetry.
In this case one is led to the following 
replacement  for
the quadratic Skyrmion Lagrangian term:
\be \ba{ll} \label{subst}
\displaystyle{ U(\br) \to \frac{\eta(\br)}{f_{\pi}} U(\br) . }
\ea
\ee 
A similar model had been considered previously \cite{7b} with
basically the same numerical results.
This replacement leads to the linear sigma model in the quadratic kinetic
 term. 
The potential term is introduced to complete the linear 
sigma model Lagrangian
in the limit in which it becomes the $\lambda \sigma^4$ model which
presents a potential with the spontaneous breakdown of symmetry.
In the fourth order Skyrme term
yields ambiguous results which are scale non-invariant Lagrangian terms and 
unstable solutions in the usually considered
physical region for the parameters of the model \cite{TESEME,RT98+DIGHOLE}. 
Therefore it was not considered.
The value of the condensate in the vacuum is found by the minimization of 
its potential.
In the vacuum (as well as
the presence of hadron(s)) 
the chiral radius as a dynamical variable 
has already been considered in \cite{RT98+DIGHOLE,CR01,massapion,7a}.
In the spatially homogeneous case of $\eta (r) = f_{\pi} $ (constant),
 the usual 
Skyrmion equation and solutions are obtained
in the infinite sigma mass limit. 
A complete account of solutions for these equations
will be shown elsewhere.
The boundary conditions for the respective 
Euler Lagrange equations are discussed
 in \cite{RT98+DIGHOLE}.
The ``quanta'' of $\eta$ field may 
be expected to correspond to the lightest 
scalar-isoscalar meson, the sigma. 
It seems to be a very broad
isoscalar resonance in the s-wave of low energy $\pi-\pi$ scattering and 
other processes,
with a mass in the range $500-800$ MeV  
\cite{TORNQVIST,BEDIAGA,TESEME,RT98+DIGHOLE}.

\subsection{ Magnetic Moments} 

The magnetic moments ($\mu_i$) of the proton and neutron are
calculated as the averaged value of the operator:
\be \ba{ll} \label{magmom}
\displaystyle{ \mu_i = 
\epsilon_{ijk}
r^j \left( \frac{1}{2} B^k + J_{V^{(3)}}^k \right),
}
\ea
\ee
where $B^k$ and $J_{V^{(3)}}^k$ are the isoscalar baryonic current 
and the third component of the isovector current respectively.
In the usual Skyrmion model \cite{ADKINS} 
the proton and neutron $\mu$ yields respectively:
\be \ba{ll} \label{mag1}
\displaystyle{ \mu_p = 2 M_N \left\{ \frac{1}{12 {\cal I}} <r^2>_{E,0} 
+ \frac{{\cal I}}{6} \right\} ,
} \\
\displaystyle{ \mu_n = 2 M_N \left\{ \frac{1}{12 {\cal I}} <r^2>_{E,0} 
- \frac{{\cal I}}{6} \right\} ,
}
\ea
\ee
where: $M_N$ is the nucleon mass - either the observable or the
Skyrmion one - and:
\be \ba{ll} \label{}
\displaystyle{ {\cal I} = 4 \pi \int d r r^2 \frac{s^2}{6} 
\left\{ 4 \eta(r)^2 + \frac{4}{e^2} \left( F'^2 + \frac{s^2}{r^2}
\right) \right\}
}\\
\displaystyle{ <r^2>_{E,0} = - \frac{2}{\pi} \int d r r^2 s F'
,}
\ea
\ee
which are the moment of inertia of the (quantized) rotating soliton 
and the averaged squared isoscalar electric radius respectively 
\cite{ADKINS}.

\subsection{ Numerical results and discussion}

For several values of the pion mass 
($m_{\pi} = 0, 139, 200, 400, 600, 700, 800$ MeV), 
the magnetic moments  of the neutron and of the proton were calculated for 
both Skyrmion models discussed above. 
For the extended Skyrmion model (Skyrmignon, with the 
scalar field) two calculations were done with expressions
(\ref{magmom}): 1) with the observed nucleon mass
(nearly $M_N \simeq 940$ MeV) and 2) with the mass resulting from the 
topological soliton numerical solution \cite{ADKINS,RT98+DIGHOLE}.

We  compare the results for the magnetic moments with different 
pion masses obtained from the calculations with
the Skyrmion and extended Skyrmion models  with those obtained from
1)the cloudy bag model \cite{LLT} and 2) 
hadronic model based on chiral perturbation theory (CPT) \cite{HEMMERTWEISE}
which are used for extrapolations in Lattice calculations \cite{LLT}.
Several resulting points and curves are shown and compared 
in Figure 1.

The parameterization found out and proposed in \cite{LLT}
for the dependence of the magnetic moment on the pion mass 
using the Cloudy Bag Model is given by:
\be \ba{ll} \label{magmom-extr}
\displaystyle{ \mu_i = \frac{A_i}{1 + B_i m_{\pi} + C_i m_{\pi}^2}
,}
\ea
\ee
where $i = n, p$ distinguishes the neutron from the proton, $A, B, C$ are
coefficients found by fitting the points from calculations shown in 
figure 1. 

The results obtained with the Skyrmion and the extended Skyrmion (
coupled to the sigma)
models are similar to those
of the CBM and CPT.
This may bring a certain credibility to both results but there is no
an unique desirable possible extrapolation which therefore depends on the
physical input of the effective model.
In particular we notice that although the behavior (slopes) of all 
the curves are similar, 
for realistic pion mass the resulting values for the magnetic moment
may be  quite different, mainly of the proton.
All the calculations done, shown in the figure 1, yield points which
can be nearly fitted with  expression (\ref{magmom-extr}).
However the coefficients are somewhat different for each calculation and
there seems to exist other extrapolations which may fit the points.

{\bf \Large \bf Acknowledgements}

F.L.B. was supported by FAPESP, Brazil. 
The author also thanks M.R. Robilotta for several discussions about 
the Skyrmion model and M. Nielsen for suggesting the calculation
of large pion/quark masses.

\vspace{1cm}

{\bf Figure 1.}

{\it Magnetic moments of the neutron and of the proton as functions of the 
squared pion mass (GeV$^2$). Full (dotted) line for the proton (neutron)
$\mu$ proposed by L.L.T. \cite{LLT}; plus + (times $\times$) obtained 
from the usual Skyrmion model for $\mu_n$ ($\mu_p$); 
dashed (short-long dashed)
lines obtained from the extended Skyrmion model 
\cite{RT98+DIGHOLE} using 
$M_N = 940$ MeV; up triangles $\Delta$ (down triangles $\nabla$)
obtained from the extended Skyrmignon  model 
\cite{RT98+DIGHOLE} using the nucleon
mass from the model itself, ($M_N \simeq 1250 $ MeV).
Thick dashed and solid lines are the fit found with chiral 
perturbation theory at one loop \cite{HEMMERTWEISE}.
 }

\vspace{2cm}

\begin{figure}[htb]
\resizebox{12cm}{!}{\rotatebox{270}{\includegraphics{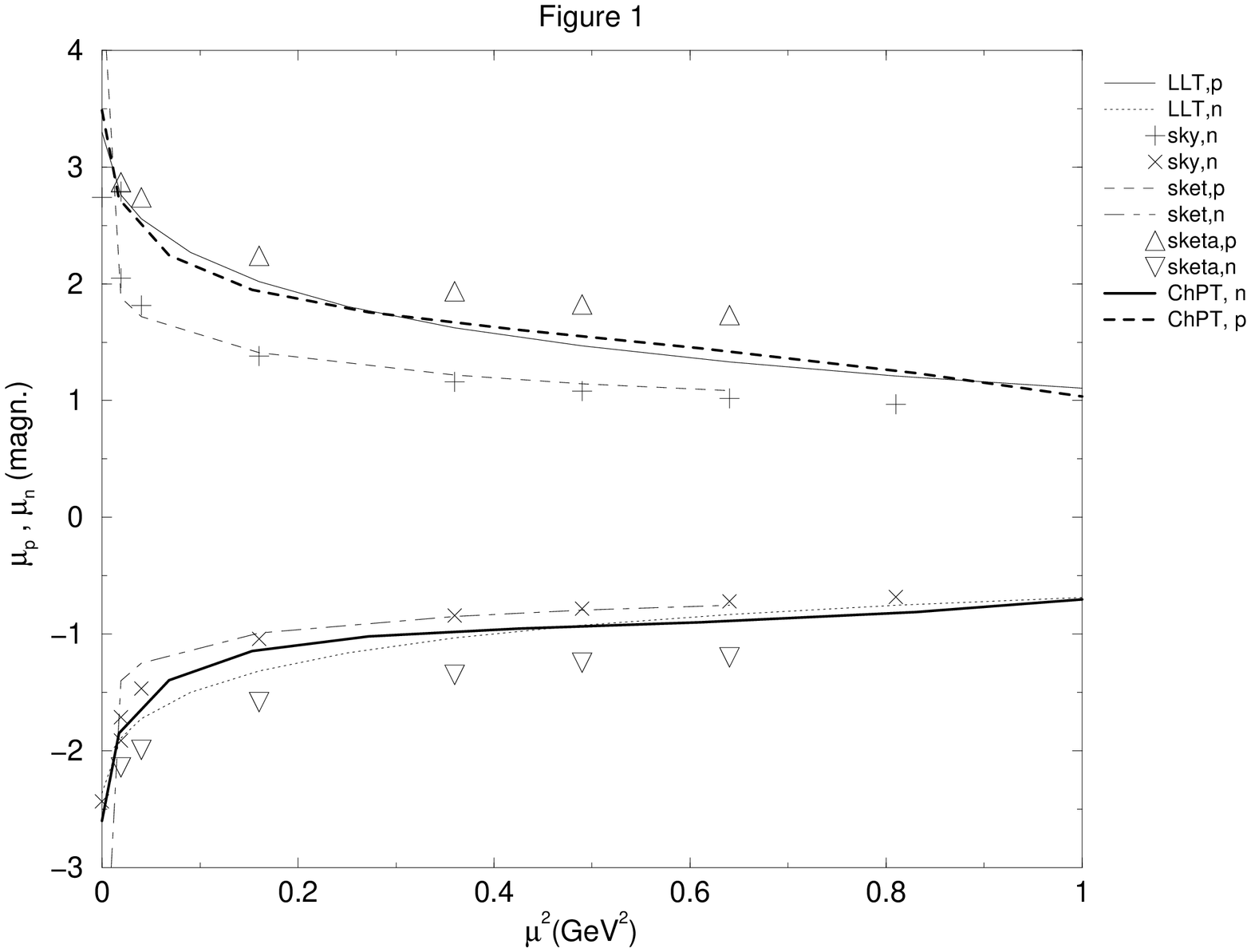}}}
\end{figure}

\end{document}